\documentclass[pra,twocolumn,superscriptaddress]{revtex4-1}
\usepackage{graphicx}
\usepackage{float}
\usepackage{mathtools}
\usepackage{amsmath}
\usepackage{amssymb}
\usepackage[colorlinks,linkcolor=blue,anchorcolor=blue,citecolor=blue,urlcolor=blue]{hyperref}

\begin{document}

\title{Onset of ferromagnetism for strongly correlated electrons in one-dimensional chains}

\author{Hernan B. Xavier}
\affiliation{Department of Experimental and Theoretical Physics - UFRN, Natal, Brazil}
\author{Evgenii Kochetov}
\affiliation{Bogoliubov Laboratory of Theoretical Physics, Joint
Institute for Nuclear Research, Dubna, Russia}
\author{Alvaro Ferraz}
\affiliation{Department of Experimental and Theoretical Physics - UFRN, Natal, Brazil}
\affiliation{International Institute of Physics - UFRN, Natal, Brazil}

\begin{abstract}
The existence of the Nagaoka ferromagnetism is examined in the context
of the one-dimensional $U=\infty$ Hubbard model. We construct the exact quantum
partition function to describe the physics of such a regime. Our calculation reveals that, while the ground state in an open chain
is always spin-degenerate, in a finite size closed chain with at least one vacancy, the ground state
can only be ferromagnetic when the number of electrons is less
or equal to three. Our results shed more light on a very recent experimental verification of Nagaoka ferromagnetism in a
quantum dot set up. 
\end{abstract}

\maketitle

\section{Introduction}

The interplay between strong correlations and the onset of ferromagnetism in itinerant electronic systems has long intrigued the many-body physics community.
In this respect, a rigorous result obtained from a single band Hubbard model (HM) in the early 1960s by Nagaoka stands as one of the most prominent theoretical landmarks \cite{Nagaoka1965-1966}.
In a nutshell, the Nagaoka theorem (NT) establishes that, for certain lattices, in the infinitely coupled regime, the presence of a single vacancy in the almost half-filled system yields a totally polarized ferromagnetic (FM) ground state.
Unfortunately, until very recently, the experimental verification of itinerant magnetism in such a regime seemed unattainable.
Despite the great success of the quantum simulations of the HM in cold atoms, no observation of such a FM ground state has been reported so far in those systems \cite{Bloch2012}.
Conversely, semiconductor based quantum dot arrays are systems that have also attracted a lot of interest as viable alternatives to realize experimentally the physics of the HM \cite{Hensgens2017}.
As a result, it was not surprising that the first experimental verification of the Nagaoka result was produced in a small scale quantum dot set up \cite{Dehollain2019}.
This new experiment, especially prepared for such a purpose, consists of a four-site quantum dot plaquette filled with three electrons.

In such small quantum systems, 
it is perfectably possible to approach this problem by means
of an exact
diagonalization of the Hamiltonian
for a fixed number of electrons. 
Indeed, by taking into account both 
distant-neighbor hopping and
Coulomb couplings, 
another theoretical work has provided further evidence that
the FM ground state of three electrons in the four-site plaquette
is robust in the presence of long range
Coulomb interactions \cite{Buterakos2019}.
In addition to this they also consider the case of four electrons in a five-site ring, which no longer displays Nagaoka ferromagnetism.
However, notwithstanding the simplicity of the HM, 
we are still not able to make general claims about what happens to the Nagaoka result if the Hubbard $U$ coupling is reduced from
its infinite value, or if the number of vacancies is kept finite in the thermodynamic regime \cite{Ivantsov2017}.
On top of that,
it is well known that most of the conventional
mean-field approximations
and perturbation schemes are both ineffective to deal
with such a strongly correlated regime. 
Particularly in one-dimension, the NT is not directly
applicable,
and for a long time it was further believed that
such a FM ground state might even contradict the Lieb-Mattis theorem (LMT) \cite{Lieb1962}.
However this is not the case. In view of the fact
that the LMT only applies to open chains, the existence
of itinerant ferromagnetism in closed chains was left entirely open until now.

In this article, we present an alternative calculation for the exact quantum partition function of the $U=\infty$ HM in one-dimensional chains of finite size.
Since our analysis already encompasses scenarios for different
numbers of electrons and lattice sizes, we are able to make precise statements about the condition for the onset of itinerant ferromagnetism in closed chains.
Our results are in full agreement with the experimental observations reported by Dehollain \textit{et al}. \cite{Dehollain2019}. Moreover we make new predictions which can also be tested experimentally.

\section{The model}

We will consider the infinite coupling regime of the
HM in one spatial dimension. To begin with, the HM describes, otherwise,
free band electrons interacting via a on-site repulsive interaction of strength $U$ \cite{Hubbard1963}.
For a single conduction band, the corresponding HM Hamiltonian reads
\begin{equation}
H_{\text{HM}}=-\sum_{ij}t_{ij}c_{i\sigma}^{\dagger}c_{j\sigma}+U\sum_{i}n_{i\uparrow}n_{i\downarrow}.\label{eq:HM-hamiltonian}
\end{equation}
where $c_{i\sigma}^{\dagger}$ ($c_{i\sigma}$) is the fermionic operator
that creates (annihilates) an electron on the lattice site $i$ with
spin projection $\sigma=\uparrow,\downarrow$; and the operator $n_{i\sigma}=c_{i\sigma}^{\dagger}c_{i\sigma}$
stands for the on-site spin-$\sigma$ electron number.
The quantum dynamics of the Hamiltonian (\ref{eq:HM-hamiltonian})
preserves the numbers of spin-up and spin-down electrons separately
since $[H,N_{\sigma}]=0$, where $N_{\sigma}=\sum_{i}n_{i\sigma}$.
This implies that both the total electron number $N=N_{\uparrow}+N_{\downarrow}$
and the total spin projection $J^{z}=\tfrac{1}{2}\left(N_{\uparrow}-N_{\downarrow}\right)$
are conserved in the system. Henceforth, the eigenenergies of (\ref{eq:HM-hamiltonian})
can always be labeled by the quantum numbers $N$ and $J_{z}$,
or, equivalently, by $N_{\uparrow}$ and $N_{\downarrow}$. We restrict ourselves to the so called hole-doped scenario with $N\leq L$, $L$ being the total number of sites. 
This can be done without loss
of generality since, by performing an appropriate particle-hole transformation in (\ref{eq:HM-hamiltonian}),
one can always recover the electron-doped energy solutions as well.

Now we turn to the large coupling limit of the HM. If the $U$ coupling
becomes the dominant energy scale in the system, the doubly occupied
electron states immediately fall into disfavour. Indeed, in the $U=\infty$
limit, they are removed altogether from the set of available
on-site states, and the hole-doped HM Hamiltonian (\ref{eq:HM-hamiltonian})
is then reduced to the projected hopping term
\begin{equation}
H=-t\sum_{\langle ij\rangle}X_{i}^{\sigma0}X_{j}^{0\sigma}.\label{eq:U-infinity-HM-hamiltonian}
\end{equation}
Here $X^{ab}=\left|a\right\rangle \left\langle b\right|$ with $\{\left|a\right\rangle \}=\{\left|\downarrow\right\rangle ,\left|\uparrow\right\rangle ,\left|0\right\rangle \}$
are the standard on-site Hubbard operators \cite{Wiegmann1988}. For simplicity, the hopping amplitudes $t_{ij}$ are
assumed to be non zero only for nearest neighbor sites $i$ and $j$.

\section{Partition function}

We work in the grand-canonical ensemble, and the quantum partition function associated with
the Hamiltonian (\ref{eq:U-infinity-HM-hamiltonian}) is 
given by
\begin{equation}
\mathcal{Z}=\text{tr }e^{-\beta \left(H-\mu \sum_i X^{00}_{i}\right)}.\label{eq:partition-function}
\end{equation}
Here $\beta$ is the inverse temperature, and the trace must be taken over a complete set of projected states, such as the $su(2|1)$ space states \cite{Ferraz2011}.
Notice that a chemical potential $\mu$ was
introduced to keep track of the number of vacancies. Naturally, this
number is always equal to the difference between the number of sites
$L$ and the total number of electrons $N$. 

\subsection{The open chain}

In an open chain, the spectrum of the Hamiltonian (\ref{eq:U-infinity-HM-hamiltonian})
is completely degenerate with respect to the spin configurations, and
the partition function can be evaluated with relative ease. The physical
intuition behind this degeneracy is quite simple to understand.
The existence of boundaries, in addition to the impossibility of exchanging
their relative ordering, automatically prevents the projected electrons to access
different spin configurations. For example, although the states $\left|\downarrow00\uparrow\right\rangle $
and $\left|\uparrow00\downarrow\right\rangle $, representing two electron states in four sites, belong to the same
$N$ and $J^{z}$ subspaces, they are dynamically inaccessible to
each other. Consequently, for each spin configuration, these projected
electrons behave essentially as spinless fermions. Hence, the resultant
partition function for the open chain is just
\begin{equation}
\mathcal{Z}^{o}(\beta,z)=\prod_{p}^{\Omega}\left(z+2e^{-\beta t_{p}}\right),\label{eq:partition-function-OBC}
\end{equation}
where $z=e^{\beta\mu}$ is the fugacity, $t_{p}=-2t\cos p$ is the
electron dispersion, and $\Omega$ is a set of momenta defined as $\Omega=\{p:p_{n}=\frac{\pi n}{L+1},n=1,\dots,L\}$.
It is worthwhile to notice that, when we turn off the hopping ($t=0$),
formula (\ref{eq:partition-function-OBC}) reduces straightforwardly to
$\mathcal{Z}=\left(2+z\right)^{L}$. This is a key property, since, if we further remove the chemical potential by taking 
$z=1$, one is able to recover the appropriate number of degrees of freedom associated with
the projected Hamiltonian (\ref{eq:U-infinity-HM-hamiltonian}).

From the partition function formula (\ref{eq:partition-function-OBC})
one can have access, not only to the full spectra of the system, but
also to some interesting finite temperature effects. For instance,
if one computes the occupation number as a function of the chemical
potential and temperature, the result is no longer the conventional Fermi-Dirac
distribution. This indicates that, despite its simplicity, the system
never ceases to have a strongly correlated nature. Nevertheless, in the zero temperature regime ($\beta\rightarrow\infty$),
$\mathcal{Z}^{o}(\beta,z)$ has a quite simple asymptotic behavior
and the expression $E^{\text{GS}} =-\lim_{\beta\rightarrow\infty}\left(\frac{\partial}{\partial\beta}\ln\mathcal{Z}^{o}(\beta,z)\right)_{z}$ for the corresponding ground-state energy reduces to
\begin{align}
E^{\text{GS}}=t-t\csc\left[\tfrac{\pi}{L+1}\left(\tfrac{1}{2}\right)\right]\sin\left[\tfrac{\pi}{L+1}\left(\tfrac{2N+1}{2}\right)\right].
\end{align}
Moreover, in the thermodynamic limit ($L\rightarrow\infty$,
$N/L\rightarrow n_{e}$), the formula above assumes the form 
\begin{equation}
\frac{E^{\text{GS}}}{L}=-\frac{2t}{\pi}\sin(\pi n_{e}),\quad n_{e}\leq1,\label{eq:HD-GS-energy-thermodynamic}
\end{equation}
which coincides with the exact result obtained earlier by Ogata and Shiba \cite{Ogata1990}
making use of the Bethe ansatz.

\subsection{The closed chain}

In a closed chain the situation is different. Without the open boundaries
to restrain the moving particles, they become free to jump around the loop
and to permute cyclically their spin positions. Certainly, not all
spin configurations are equivalent to each other and, in view of that,
the spin degeneracy is partially lifted. While a fully polarized spin
state, e.g., $\left|\uparrow\uparrow00\uparrow\right\rangle $, only
has translational degrees of freedom, which can be related solely
to the number of vacancy positions, the same does not generally
hold to other spin states. One example of that is the state
$\left|\uparrow\uparrow00\downarrow\right\rangle $, which can also
access dynamically all the other states which are cyclic permutations
of these spins, e.g., $\left|\uparrow\downarrow00\uparrow\right\rangle $
and $\left|\downarrow\uparrow00\uparrow\right\rangle$.
Thanks to this feature, the HM in a closed chain can be considered as an example of a \emph{quantum necklace}. Indeed, the number of dynamically disconnected
subspaces $D_{s}(N)$ in this model is equal to the number of distinct
necklaces that can be made with the $N$ projected fermions of spin-$s$. Using
the Burnside's lemma \cite{PolyaBook}, it turns out that
\begin{equation}
D_{s}(N)=\frac{1}{N}\sum_{d|N}\varphi(\tfrac{N}{d})\left(2s+1\right)^{d},\label{eq:necklace-number}
\end{equation}
where $\varphi(x)$ is the Euler's totient function, which is defined as the
number of positive integers between $1$ and $x$ that are coprime
to $x$. Here $d|N$ stands for a sum over the natural divisors of
$N$. In particular, the first few values of $D_{s}(N)$ for
projected electrons are
\begin{equation}
\begin{split}
D_{1/2}(1) =2,\qquad D_{1/2}(2)&=3,\qquad D_{1/2}(3)=4, \\
D_{1/2}(4) =6,\qquad D_{1/2}(5)&=8,\qquad D_{1/2}(6)=14, \\
D_{1/2}(7) =20,\qquad D_{1/2}(8)&=36,\qquad D_{1/2}(9)=60.
\end{split}\label{eq:necklace-number-electrons}
\end{equation}

To find the corresponding spectra, now we just need to diagonalize separately each one of
those distinct necklaces.
The important point is that these necklaces can be further classified according to their irreducible cyclic symmetry of the spin configuration $C_{d}$, where $d$ is a natural divisor of $N$. 
For such a $C_{d}$ necklace, the projected electron momenta are quantized in the form
\begin{equation}
p_{n}=\frac{2\pi}{L}n+\frac{2\pi}{NL}\nu,\label{eq:necklace-momenta-quantization}
\end{equation}
where $n=0,\dots,L-1$ and $\nu=0,\frac{N}{d},\dots,(d-1)\frac{N}{d}$.
Equation (\ref{eq:necklace-momenta-quantization}) is just comprised of the $\frac{2\pi n}{L}$ contribution, which results from the existing translational invariance, and the $\frac{2\pi \nu}{NL}$ contribution, which is associated with the $C_{d}$ cyclic invariance of the spin configuration.
This momentum shift is produced by the relative
movement of the ``spin background" as the vacancies move along the sites \cite{Ivantsov2019}.
Making use of all these ingredients, we are in a position
to write the canonical partition function as
\begin{equation}
Z_{N}^{c}=\sum_{\nu=0}^{N-1} M^{N}_{\nu} \sum_{p_{1}<p_{2}<\dots<p_{N}}^{\Gamma(\nu)}e^{-\beta\left(t_{p_{1}}+t_{p_{2}}+\dots+t_{p_{N}}\right)},
\label{eq:can-partition-function-CBC}
\end{equation}
where $\Gamma(\nu)$ is the momenta set defined as $\Gamma(\nu)=\{p:p_{n}=\frac{2\pi n}{L}+\frac{2\pi \nu}{NL},n=0,\dots,L-1\}$,
and the symbol $M^{N}_{\nu}$ plays the role of a degeneracy factor.
The latter arises because more than one necklace can contribute to a particular $\nu$ solution.
In special, since all the necklaces contribute to the $\nu=0$ solution, its degeneracy factor is always equal to the total number of distinct
necklaces, i.e., $M_{0}^{N}=D_{s}(N)$.
The remaining $M_{\nu}^{N}$ factors can be determined in similar manner, and depend on the particular
divisor structure of the cyclic group ${C}_{N}$.
Notably, if $N$ is a prime number,
the necklaces are only irreducibly symmetric to $C_{1}$ or $C_{N}$, and, therefore,
it turns out that $M_{0}^{N}=D_{s}(N)$,
and $M_{\nu}^{N}=D_{s}(N)-\left(2s+1\right)$ for all $\nu>0$.
In general, if $\nu$ and $\nu'$ share the same greatest
common divisor with $N$, i.e., $\gcd(N,\nu)=\gcd(N,\nu')$, their
degeneracy factors are equal as well, i.e., $M_{\nu}^{N}=M_{\nu'}^{N}$.
Some examples and explicit values of the degeneracy factors are available in the Appendix \ref{appdx:degeneracy-factor}.
Notice that, for $N=L$, as the projected electrons cannot move, the
energy $E=t_{p_{1}}+\dots+t_{p_{N}}$ in (\ref{eq:can-partition-function-CBC}) invariably vanishes, and provided $\sum_{\nu=0}^{N-1}M_{\nu}^{N}={2}^{N}$, the canonical particion function yields $Z_{N=L}^{c}=2^{L}$.

Hence, by combining the contributions for different electron numbers, we can obtain the sought-after partition function formula
\begin{equation}
\mathcal{Z}^{c}(\beta,z)=\sum_{N=0}^{L}z^{L-N}Z_{N}^{c}.\label{eq:partition-function-CBC}
\end{equation}
Here, similarly to what happens with the open chain,
$\mathcal{Z}^{c}(\beta,z)$ also reduces to $\mathcal{Z}=(2+z)^{L}$
at $t=0$. Additionally to that, it looses
its spin dependence as the system approaches the thermodynamic limit.
This takes place because, as $L\rightarrow\infty$, all the $\Gamma(\nu)$ sets
become isomorphic to $S^{1}$. In fact, this is no
surprise since the solutions for open and closed boundary conditions
must approach each other in this limit, i.e.,
\begin{equation}
\lim_{L\rightarrow\infty}\frac{\mathcal{Z}^{c}(\beta,z)}{\mathcal{Z}^{o}(\beta,z)}=1.
\end{equation}

\section{Nagaoka ferromagnetism in one-dimensional chains}

Following Nagaoka \cite{Nagaoka1965-1966}, the ground state is said to be FM only if it consists solely of the maximum total spin states, and there is no other lower total spin state which is degenerate to it.
Thus, since in the $U=\infty$ limit, the spectra is completely degenerate with
respect to spin in an open chain, its ground state is always
non-magnetic.
Naturally, in real systems, the $t/U$
corrections will lift such a degeneracy. However, the absence of
Nagaoka ferromagnetism, even in this finite coupling limit, is reassured by the LMT \cite{Lieb1962}.

In contrast with that, for a closed chain, the ground state 
can be FM if some precise conditions are met. In the first place, the number 
of electrons cannot be greater than three.
This follows from the fact that 
a genuine FM ground state, 
which is contained in our $\nu=0$ solution, 
can only be realized if $M^{N}_{0}$ is identical to
the degeneracy coming from the states with maximum spin $J^\text{max} = sN$,
i.e., if $D_{s}(N)=2sN+1$.
In fact, \textit{when the number of electrons is equal to three, if there is at least one vacancy,
the ground state of (\ref{eq:U-infinity-HM-hamiltonian})
is FM, for $t>0$, in any finite closed chain; and, for $t<0$, in any finite closed chain with an even number of sites}.
On the other hand, \textit{when the number of electrons is equal to two, if there is at least one vacancy, the ground state of (\ref{eq:U-infinity-HM-hamiltonian}) is FM only for the case in which $t<0$, and the finite closed chain has an odd number of sites}.
Consequently, for $t>0$, a closed chain of size $L=5$
with two vacancies exhibits a FM ground state, although the same
chain has a non-magnetic ground state for the single vacancy case.
The key point is that, differently from the standard NT condition
which is characterized by the presence of a single vacancy, for closed
one-dimensional chains, the existence or not of such a FM ground state is
directly determined by the total number of electrons in the system. 
The present result can be considered as a natural extension of the NT for one dimensional chains.
As a matter of fact, these two concepts
are complementary to each other and, for the $L=4$ case, which can be interpreted
either as a $2\times2$ array or a closed chain, they equally
predict a FM ground state if only a single vacancy is present in the system.
Further details about the role of the spin can
be found in the Appendix \ref{appdx:absenceFM-GS-higher-spin}.

\begin{figure}
\begin{centering}
\includegraphics[scale=0.42]{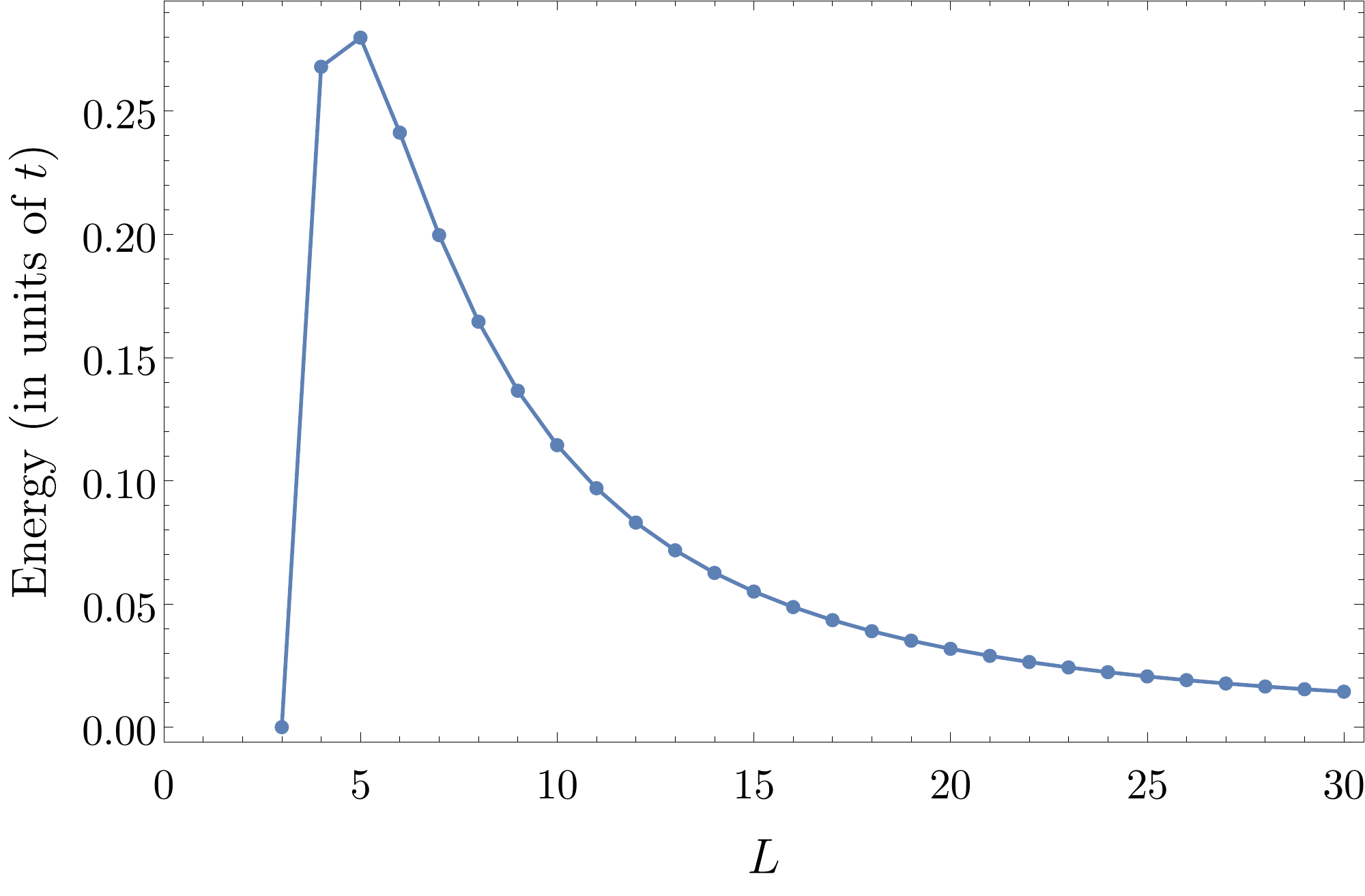}
\par\end{centering}
\caption{Energy gap between the FM ground state and the first unpolarized excited-state. For $L=N=3$, all spin states are degenerate and have energy equal to zero.
\label{fig:gap-3/2-1/2}}
\end{figure}

However, the feasibility to observe such phenomenon in closed chains will
certainly depend on the precision of the instruments to measure the energy difference
between the FM ground state and the first unpolarized excited-state.
In particular, for $N=3$ electrons, for $t>0$, the expected energy gap between
the lowest $J=\tfrac{3}{2}$ and $J=\tfrac{1}{2}$ spin states is
\begin{equation}
E_{1/2}-E_{3/2}=4t\sin^{2}(\tfrac{\pi}{3L})\left[1+2\cos(\tfrac{2\pi}{L})\right].\label{eq:energy-gap}
\end{equation}
In Figure \ref{fig:gap-3/2-1/2}, we plot the values of this energy
gap as a function of the chain size $L$, for $3\leq L\leq30$. 
As displayed there, the energy gap decreases as $L$ increases and eventually approaches zero in the $L\rightarrow\infty$ limit.
This is a clear indication that there is no FM ground state in the thermodynamic limit.
Thus such
a FM state is most likely to be detected in chains with a small number of sites.
In special, since the $L=4$ case was already verified experimentally \cite{Dehollain2019}, the $L=5$ case, which produces the maximal energy gap for $N=3$ electrons can, in principle, be readily detected by similar experimental set ups.

\section{Discussion}

We study the quantum dynamics of the Hubbard model at infinite coupling in one-dimension. Using
the fact that the sequence of spins cannot be changed by
hopping vacancies, we derive an explicit expression for the associated partition
function.
From there, we deduce a precise condition for the onset of kinetic ferromagnetism in closed chains, while having free access to excited states and to compute energy gaps.

As a matter of fact, our approach is applicable to other one-dimensional contrained hopping models, such as for spinful hard-core bosons \cite{Girardeau1960,Paredes2004,Stoferle2004} and the strongly coupled heavy-fermion systems \cite{Sigrist1992,Pepino2008}. 
However, the question of what happens in the presence of projected hopping with a more extended range is far from settled \cite{Buterakos2019}. The inclusion of more non-vanishing hopping matrix elements, increases the connectivity of the spin configurations, and may lead to some rather unexpected
non-trivial results, such as the emergence of metastable FM
states \cite{Ivantsov2019} or even to kinetic antiferromagnetic order on frustrated lattices \cite{Haerter2005}.

We conclude with some remarks and a speculation concerning the role of dimensionality. 
It is well-known that Nagaoka ferromagnetism requires the existence of closed loops \cite{Nagaoka1965-1966}. 
Moreover, the lattice should satisfy a connectivity condition: the elementary loops should pass through no more than four sites \cite{Tasaki1989}.
Since this is a condition derived for the single vacancy scenario, it implies that such a loop cannot contain more than three electrons while the vacancy travels by. 
This idea may suggest that the kinetic ferromagnetism emerges in a higher dimensional lattice as a sort of a ``confined phase," i.e., a phase in which its physical properties are determined by
the presence of such elementary loops.
Such a description is realizable, at least in the presence 
of a nearest-neighbor attraction \cite{Kornilovitch2014}. However the stability of this phenomenon in more general settings is still unknown.

\acknowledgments

This work was supported by the CAPES agency -- Brazil -- Finance Code 001. One of us (A.F.) also acknowledges the financial support from the Ministry of Education (MEC) and from the CNPq agency -- Brazil.

\appendix

\section{Necklaces and degeneracy factors}\label{appdx:degeneracy-factor}

\begin{figure}
\begin{centering}
\vspace{.5cm}
\includegraphics[scale=0.45]{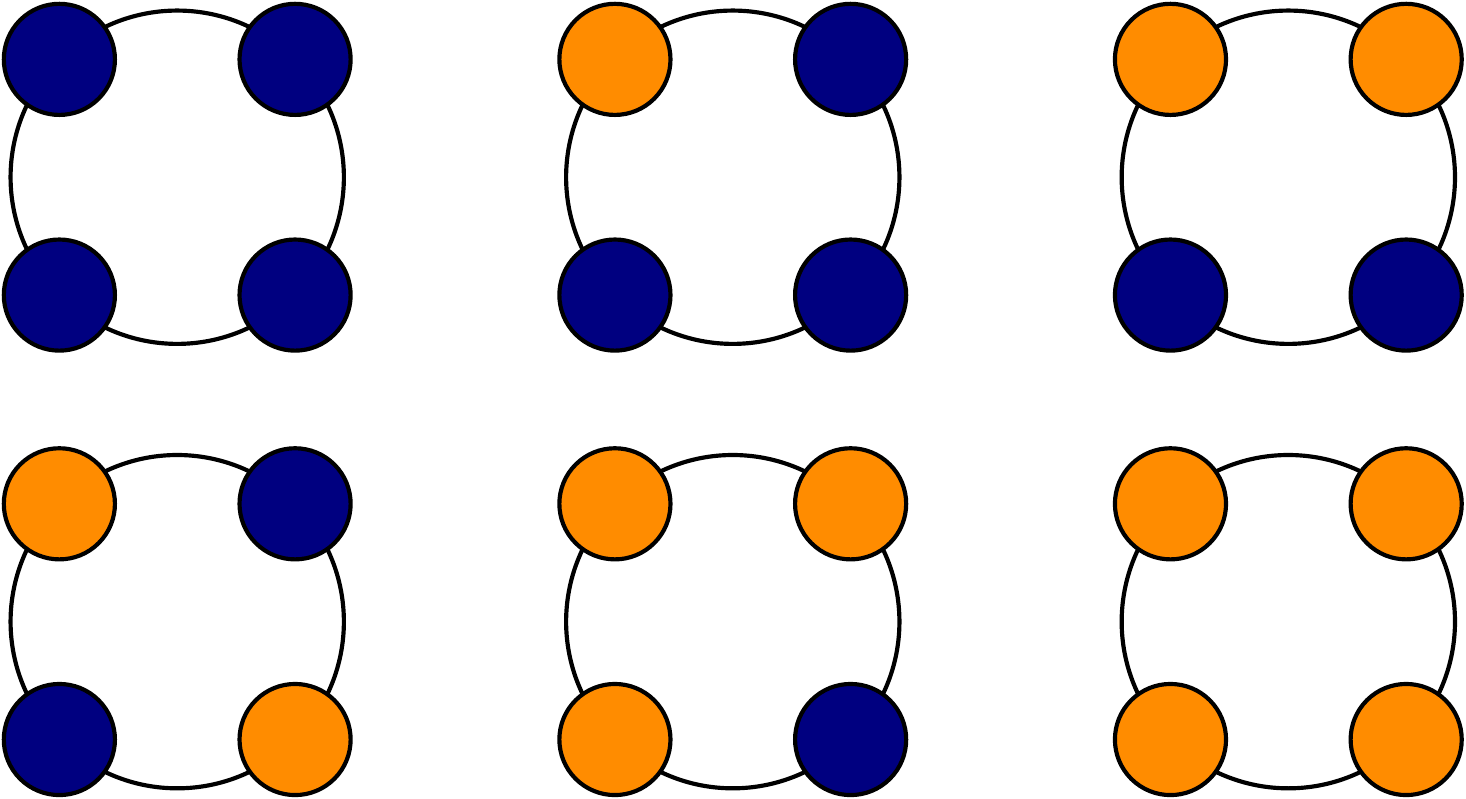}
\par\end{centering}
\caption{Necklaces that can be formed with $N=4$ projected
electrons. Each color represents one of the two spin polarization
states.\label{fig:4-2-necklace}}
\end{figure}

In this appendix, we provide a couple of simple examples to show in more detail how to determine explicitly the $M_{\nu}^{N}$ factors.
First, we consider the scenario with just $N=4$ electrons.
Due to its divisor structure, in addition to $C_{4}$, there are also necklaces whose irreducible
cyclic symmetry are $C_{1}$ or $C_{2}$.
For clarity, we display the six distinct necklaces that can be formed with these electrons in Figure \ref{fig:4-2-necklace}.
As one can readily identify, two of those necklaces are irreducibly symmetric to
$C_{1}$, and just one of them, the fourth one in Figure
\ref{fig:4-2-necklace}, is irreducibly symmetric to $C_{2}$. Thus, from the identities given in the main text, we get that $M_{0}^{4}=6$,
$M_{1}^{4}=M_{3}^{4}=3$ and $M_{2}^{4}=4$. If we now move on to the case
with $N=8$ electrons, the relevant cyclic groups are $C_{8}$, $C_{4}$,
$C_{2}$ and $C_{1}$. However, in spite of the change in the number of electrons, the number of necklaces irreducibly symmetric related to a particular $C_{d}$ follow the same pattern as before.
Namely, there are still only two necklaces irreducibly symmetric to $C_{1}$,
one to $C_{2}$, and three to $C_{4}$. 
Therefore $M_{0}^{8}=36$, $M_{1}^{8}=M_{3}^{8}=M_{5}^{8}=M_{7}^{8}=30$,
$M_{2}^{8}=M_{6}^{8}=33$ and $M_{4}^{8}=34$. Following this prescription,
we write down the $M_{\nu}^{N}$ factors for the first few values of $N$.
\begin{widetext}
\begin{equation}
\begin{split}N=0:\qquad &M_0^0=1;\\
N=1:\qquad  & M_0^1=2;\\
N=2:\qquad & M_0^2=3,\qquad M_1^2=1;\\
N=3:\qquad & M_0^3=4,\qquad M_1^3=M_2^3=2;\\
N=4:\qquad & M_0^4=6,\qquad M_1^4=M_3^4=3,\qquad M_2^4=4;\\
N=5:\qquad & M_0^5=8,\qquad M_1^5=M_2^5=M_3^5=M_4^5=6;\\
N=6:\qquad & M_0^6=14,\qquad M_1^6=M_5^6=9,\qquad M_2^6=M_4^6=11,\qquad M_3^6=10;\\
N=7:\qquad & M_0^7=20,\qquad M_1^7=M_2^7=M_3^7=M_4^7=M_5^7=M_6^7=18;\\
N=8:\qquad & M_{0}^{8}=36,\qquad M_{1}^{8}=M_{3}^{8}=M_{5}^{8}=M_{7}^{8}=30,\qquad M_{2}^{8}=M_{6}^{8}=33,\qquad M_{4}^{8}=34;\\
N=9:\qquad & M_{0}^{9}=60,\qquad M_{1}^{9}=M_{2}^{9}=M_{4}^{9}=M_{5}^{9}=M_{7}^{9}=M_{8}^{9}=56,\qquad M_{3}^{9}=M_{6}^{9}=58.
\end{split}
\end{equation}
\end{widetext}

\section{Absence of Nagaoka ferromagnetism for spin-$s>1/2$ fermions}\label{appdx:absenceFM-GS-higher-spin}

In the main text, we have used the condition $D_{s}(N)=2sN+1$ to constrain the maximum number of electrons in a Nagaoka ground state. However, it does not take much to see that, if the projected particle has a higher
value for the spin, i.e., if  $s>1/2$, the same equation only provides physical solutions for $N = 1$. Some values of 
$D_{s}(N)$
for $s>1/2$ are given below:
\begin{equation}
\begin{split}D_{1}(1)=3,\quad D_{1}(2)&=6,\quad D_{1}(3)=11;\\
D_{3/2}(1)=4,\quad D_{3/2}(2)&=10,\quad D_{3/2}(3)=24;\\
D_{2}(1)=5,\quad D_{2}(2)&=15,\quad D_{2}(3)=45.\\
\end{split}
\end{equation}

\end{document}